\documentclass[preprint,aps, preprintnumbers, nofootinbib]{revtex4}
\usepackage[dvips]{graphics}

\begin{document}

\title{Renormalization-group running cosmologies and the generalized second
law}

\author{R. Horvat\footnote{horvat@lei3.irb.hr}}
\affiliation{\footnotesize Rudjer Bo\v{s}kovi\'{c} Institute,
         P.O.B. 180, 10002 Zagreb, Croatia}

\begin{abstract}

We explore some thermodynamical consequences of accelerated universes driven by
a running cosmological constant (CC) from the renormalization group (RG).
Application of the generalized second law (GSL) of gravitational
thermodynamics to a framework where the running of the CC goes at the expense
of energy transfer between vacuum and matter, strongly restricts the mass
spectrum of a (hypothetical) theory controlling the CC running. We find that
quantum effects driving the running of the CC should be dominated by a 
trans-planckian mass field, in marked contrast with the GUT-scale upper mass bound
obtained by analyzing density perturbations for the running CC. The model
shows compliance with the holographic principle.   

\end{abstract}

\newpage

\maketitle

In ordinary quantum 
field theory (QFT) the CC is viewed as a parameter subject to RG
running and therefore is expected to run with the RG scale, usually
identified with an expansion quantity evolving smoothly enough to comply with
cosmological data. In such theories, therefore, even a `true' CC cannot be
fixed to any definite constant (including zero) owing to RG running effects.
In \cite{1}, the variation of the CC arises solely from the particle field
fluctuations, while \cite{2} represents a complementary approach in which RG
running is due to non-perturbative quantum gravity
effects and a hypothesis of the existence of an IR attractive RG fixed
point. The main theoretical obstacle to treating the CC within QFT in a  curved 
background (the most appropriate tool for studying the problem \cite{1}) is 
that a derivation of the form of decoupling of heavy-particle species in
quantum effects governing the CC running, cannot be obtained in 
a rigorous way within this framework \cite{3}. If one insists on the
familiar 
quadratic form of decoupling for heavy-matter fields 
at low energies, then one ends up with a
somewhat surprising outcome that more massive fields do play a dominant role
in the running at any scale \cite{4}. Consequently, the running in this case
becomes stronger than logarithmic, thus providing a 
viable mechanism for efficient
relaxing  of the CC  from a large value in the early universe to its tiny value
observed today. The scenario can therefore shed some light on the
hard-pressing CC problem \cite{5}. 

In addition, the above scenario for the CC running, with the choice for the
RG scale $\mu = H$, taken together with the conservation law controlling the
continuous transfer of energy between the CC and matter, may provide a
viable cosmological model of dark energy of the universe \cite{6}. Indeed, the CC
variation law, $d\rho_{\Lambda }/dz \propto
dH^2/dz$ \cite{1}, gives the CC scaling in the form
\begin{equation}
\rho_{\Lambda } = C_0 + C_2 H^2 \;,
\end{equation}
thus having a natural appearance of a nonzero constant $C_0$, while  another
(dimensionful) constant 
$C_2 $ represents the effective cumulative mass squared of an underlying QFT
and therefore is dominated by the heaviest masses. $C_0$ represents
the ground state of the vacuum, and, of course, cannot be unambiguously
set in any theory. The right amount of dark energy at present is
obtained for $C_2 \sim M_{Pl}^2 $, if $C_0 $ is subdominant in (1). However, the
presence of (even if tiny) $C_0 $ is essential here for phenomenological
reasons, since otherwise the scenario is incompatible with a transition from
a decelerated to an accelerated era for a spatially flat universe \cite{7}.
This is because the matter energy density $\rho_m $ scales with the
expansion of the universe in the same way as the variable part in (1) (see
below). The parameter $C_2 $ from the variable part of (1) could be
potentially observable in the future supernova data \cite{8}. The
strongest upper bound on $C_2 $ has been obtained recently \cite{9}, through the
numerical analysis of  density perturbations for the running CC
$((C_2)^{1/2} \sim  10^ {-3} M_{Pl})$ \footnote{Somewhat less stringent
bound was obtained in a less rigorous way in \cite{10}}. Although these
bounds on $C_2 $ mean that an interaction between matter and the CC is small
enough that (1) is dominated by the constant $C_0 $ today, the
phenomenological aspect of the model can still be considered viable.   
   
Although the above scenario looks promising in touching both the CC problem and
the problem of  
dark energy of the universe (amongst other advantages because no 
quintessence-like
scalar fields are involved whatsoever), an attempt to bring the above model in
accordance with holographic cosmology leads unavoidably to undesired
phenomenological implications \cite{11}. In \cite{11},  
a serious drawback was noticed for 
the model (1), in conjunction with the corresponding equation of 
continuity, when
trying to accommodate the prediction from the important concept of
holographic dark energy (HDE) \cite{12, 13}. Namely, such accommodation 
unavoidably sets the ground state of the vacuum to zero $(C_0 = 0)$. Thus,
as stated above, a transition between the 
two cosmological eras for flat space
(as suggested by observations as well as by inflation) cannot be obtained.
In order to bring the model (1) in agreement with the concept of HDE, a
different approach was entertained in \cite{11}, in which (1) was
investigated together with a different equation of continuity where a
transfer of energy is between the CC and a gravitational field, thus
also promoting the Newton constant to a varying quantity. In addition, the RG
scale in this case is fixed by the continuity equation, and  therefore 
cannot generally be set at $\mu = H$. Another possible remedy is a
non-saturated HDE \cite{14}.

In the present paper, we put the above model under the scrutiny of another
profound physical principle, the generalized Second Law (GSL) of
gravitational thermodynamics. In the context of modern cosmology, the
Second Law of thermodynamics  is manifest there since the initial conditions 
for cosmology have low entropy, so we can see the Second Law in operation
\cite{15}. In the problem under consideration it is adequate to invoke the
GSL because we are dealing with cosmologies in which ever accelerating
universes always possess future event horizons. The GSL
states that
the entropy of the event horizon plus the entropy of all the stuff in
the volume inside the horizon cannot decrease in time. The idea of
associating entropy with the horizon area surrounding black holes is
now extended to include all event horizons \cite{16}. We aim to restrict the
parameter $C_2 $, which controls the running of the CC, by assuming the
validity of the GSL. 

The cosmological solutions for $\rho_{\Lambda }$ and $\rho_m $, using the
equation of continuity of the form 
\begin{equation}
\dot{\rho }_{\Lambda } + \dot{\rho }_m +3H {\rho }_m = 0  \;,
\end{equation}
and the Friedmann equation for flat space read \cite{6}
\begin{eqnarray}
\rho_m & = & \rho_{m_{0}} a^{-\zeta } \;,
\end{eqnarray}
while $\rho_{\Lambda }$ is given by (1) with 
\begin{eqnarray}
C_0 & = & \rho_{{\Lambda }_0 } -\frac{3 \nu }{8 \pi } M_{Pl}^2 H_{0}^2 
\;,
\\
C_2 & = & C_0 + \frac{3 \nu }{8 \pi } M_{Pl}^2 \,.
\end{eqnarray}
A dimensionless parameter $\nu $ is defined as  
\begin{equation}
\nu =\frac{\sigma M^2 }{12 \pi M_{Pl}^2 } \;,
\end{equation}
where $M$ represents an additive mass contribution of all virtual massive
particles, $\sigma = \pm 1$ depending on whether the highest-mass particle is
a boson/fermion, $\zeta = 3(1 - \nu )$, and the subscript `0' denotes the
present-day value. Here $|\nu | \sim 10^{-2}$ means the existence of
a particle  with Planck mass (or the existence of somewhat less massive
particles with large multiplicities), $|\nu | \sim 1$ means the
existence of a particle with trans-planckian mass, $|\nu |  \sim
10^{-6}$ means the existence of a particle with GUT-scale mass, whereas
much smaller values for $|\nu |$ mean an approximate cancellation between
bosonic and fermionic degrees of freedom. As already mentioned, the variable
part of $\rho_{\Lambda } \sim \rho_m \sim H^2 $. 

We begin by considering the entropy of a variable CC  
inside the future event
horizon of a comoving observer. Its entropy inside the horizon can be
determined by Gibb's equation for the zero chemical potential
\begin{equation}
T_{\Lambda }dS_{\Lambda } = dE_{\Lambda } + p_{\Lambda }dV \;.
\end{equation}
Although we deal here with  a `true' CC, $p_{\Lambda }=-\rho_{\Lambda
}$ ($w_{\lambda }=-1$), we learn from (7) that, owing to its
variable character,  it still possesses a nonzero entropy,
\begin{equation}
T_{\Lambda }dS_{\Lambda } = V d\rho_{\Lambda } \;.
\end{equation}
The temperature of the CC fluid $T_{\Lambda }$ has to match  
the temperature of the future event horizon, which, in spite of the fact
that the degeneracy between the apparent and the event horizon is 
broken here, one assumes it to be of a de Sitter form of the Gibbons-Hawking 
temperature $T_{\Lambda }=(2 \pi )^{-1} H$ \cite{17, 18}. With the entropy
associated to the event horizon
\begin{equation}
S_{E} = \pi M_{Pl}^2 d_{E}^2 \;,
\end{equation}
the GSL states that \footnote{Note that the entropies of all other stuff
inside the horizon do not appear in (10). This is because $\dot{S}_{\Lambda }$
grossly overwhelms all of them. For instance, a contribution of ordinary
matter, $S_m \sim (\rho_m /m) d_{E}^3 $, is suppressed in (10) today 
with respect to the contribution from $S_{\Lambda }$ by a huge factor 
$mM_{Pl}/\rho_{m}^{1/2}$. Similarly for other components including the
entropy of Hawking particles. However, this may change in the asymptotic
region $a \rightarrow \infty $ since various components scale
differently with the scale factor \cite{19}. A treatment of relic
gravitational waves inside the horizon requires an extra care \cite{20}. For
related works, see \cite{21}}  
\begin{equation}
\dot{S}_{E} + \dot{S}_{\Lambda } \geq 0 \;,
\end{equation}
where overdots represent time derivatives and 
the future event horizon is given by 
\begin{equation}
d_{E} = a \int_{a}^{\infty } \frac{da}{a^2 H} \;,
\end{equation}
with $a$ being a scale factor. Upon incorporating the expression (5) for $C_2
$ into (10), we obtain the GSL requirement in a compact form
\begin{equation}
\nu \dot{H} \geq (d_{E}^{-1} \dot{)} \;.
\end{equation}
In order to make the presence of cosmic horizons manifest in the above 
scenario, we restrict ourselves to the parameter space ensuring a de Sitter
fate of the universe, i. e., $\nu < \Omega_{\Lambda }^0 \sim 0.7$. 
Note that this range also covers the interval 
of compatibility with the LSS data $-10^{-6} \leq \nu \leq
10^{-6}$ \cite{9}. In addition, to
treat the problem completely analytically, in the
following we investgate 
whether the GSL requirement (12) is fulfilled in a dark-energy
dominated phase of the expansion only.  
Thus we expand the Hubble parameter 
\begin{equation}
H^2 = \frac{8 \pi }{3} M_{Pl}^{-2} 
\left ( (\rho_{{\Lambda}_0 } -\frac{\nu }{1 - \nu }\; \rho_{{m}_0
})
+ \rho_{{m}_0 } \; \frac{1}{1 - \nu } \; a^{-\zeta } \right )
\end{equation}  
around its de Sitter value [the constant term in (13)], and keep only the
first term in order to obtain the necessary time dependence. Performing so
for $d_{E}^{-1}$ too, we finally end up with the GSL requirement expressed as a 
simple bound on
the dimensionless parameter $\nu $
\begin{equation}  
\nu (4 - 3 \nu ) \geq 1 \;.
\end{equation}
The solution of (14), considering the requirement $\nu < \Omega_{\Lambda }^0$, 
\begin{equation}
\Omega_{\Lambda }^0 > \nu \geq \frac{1}{3} \;,
\end{equation}
clearly shows that the requirement from the GSL for a de Sitter phase of the
expansion is obeyed only for a small range of trans-planckian masses. A
preferred situation \cite{22} from a view of string/M theory, in which a
positive CC at present becomes a negative one sometime in the future 
(anti de Sitter
fate of the universe), is not covered by our analysis as it corresponds to $\nu
> \Omega_{\Lambda }^0$. However, such large values for $\nu $  are
excluded anyway from recent considerations of the dynamics of
density perturbations for the running CC \cite{9}.
      
We also need to check up (for consistency) if the entropy $S_{\Lambda }$ 
obeys the bound imposed by the holographic principle. For that purpose, we
integrate (8) in the manner consistent with our expansion around the constant
term in (13). Using
\begin{equation}
\dot{H} \simeq (1 + \zeta ) (d_{E}^{-1} \dot{)} \;,
\end{equation}
we arrive at 
\begin{equation}
S_{\Lambda } = -\nu \pi (1 + \zeta ) d_{E}^{2} M_{Pl}^2 + C \;.
\end{equation}
The constant $C$ can be determined by noting that: (i) $\dot{S}_{\Lambda } <
0$, (ii) $S_{\Lambda } \geq 0$
for any sort of nonphantom dark energy, (iii) the adapted third law of
thermodynamics tells us that $S_{\Lambda } \rightarrow 0$ for $t
\rightarrow \infty $. One obtains
\begin{equation}
S_{\Lambda } = \pi \nu (1 + \zeta ) M_{Pl}^2 (d_{E_{\infty }}^{2} -
d_{E}^{2}) \;.
\end{equation}
Since $\nu (1 + \zeta ) \sim 1$ at most and $d_{E_{\infty }}^{2} -
d_{E}^{2} \ll d_{E}^{2}$ by assumption, we conclude that $S_{\Lambda }$ always
obeys the holographic bound during a de Sitter phase of the expansion. 
  
One may attempt to bring the bound on $\nu $, obtained from assuming the
validity of the GSL,  in accordance with the bound obtained from compatibility
with the LSS data by defining the CC temperature up to a constant, $b (2 \pi
)^{-1} H$, where b is a constant. This would however decrease the temperature 
of the event horizon by a factor up to $10^{6}$, which looks unacceptably 
small. In addition, as warned in \cite{18}, any value of $b$ different from 1
would admit a nonzero flow of energy between the horizon and the fluid (or
viceversa), thus destroying thermal equilibrium inherent to the FRW
geometry. 

Another attempt could be to study the law (1) together with the generalized
equation of continuity
\begin{equation}
\dot{G}_{N}(\rho_{\Lambda } + \rho_m ) + G_N \dot{\rho }_{\Lambda } +
G_N (\dot{\rho }_{m} + 3H\rho_m )  = 0  \;
\end{equation}
which opens up an extra flow of energy now at the expense of variation of 
$G_N $. To get a consistent theory with a varying $G_N $,  one normaly goes 
over some scalar-tensor theory. In the absence of a scalar-tensor theory, the 
scaling dependence of
$G_N $ from the RG entering (19), is used to perform a ``RG
improvement'' \cite{2} instead, either at the level of Einstein's equation or at the
level of the Einstein-Hilbert action. Considering the present bounds on the 
variation of the gravitational coupling \cite{23}, one is however 
skeptical how this
could change the running of the CC to such an extent as to come up with
something very different from (14). In any case, any approach of this kind 
would therefore inevitably require a choice for the RG scale at a significant 
variance with $\mu = H$. Let us mention though that an analysis of density
perturbations for the variable CC and the variable $G_N$ is not available yet.

In conclusion, we have shown how the GSL applied to a model with a variable
CC based on the RG effects from standard QFT, sets the lower bound on the
heaviest mass in the theory, which cannot be pushed below the Planck
scale. The model where the only  transfer of energy is between the CC and
matter, complies with the holographic bound albeit not with the concept of
HDE. If another channel for transfer of energy opens up
(between $\rho_{\Lambda }$ and $G_N $), a merging with the concept of HDE
density results in moving the heaviest possible masses towards the Planck
scale, while the lowest mass sets in around the quintessence-like mass scale
\cite{24}. It therefore seems that at least as far as QFT is concerned, the
predictions from the GSL and HDE go hand in hand.

{\bf Acknowledgment. } This work was supported by the Ministry of Science,
Education and Sport
of the Republic of Croatia under contract No. 098-0982887-2872.

\end{document}